# E-commerce Webpage Recommendation Scheme Base on Semantic Mining and Neural Networks

**Wenchao Zhao[1],\*, Xiaoyi Liu[2], Ruilin Xu[3], Lingxi Xiao[4], Muqing Li[5]**

[1]University of Science and Technology of China, Hefei, Anhui, China
[2]Arizona State University, Phoenix, USA
[3]The University of Chicago, Chicago, USA
[4]Georgia Institute of Technology, Atlanta, USA
[5]University of California San Diego, La Jolla, USA
*Correspondence Author

**Abstract:** *In e-commerce websites, web mining web page recommendation technology has been widely used. However, recommendation solutions often cannot meet the actual application needs of online shopping users. To address this problem, this paper proposes an e-commerce web page recommendation solution that combines semantic web mining and BP neural networks. First, the web logs of user searches are processed and 5 features are extracted: content priority, time consumption priority, online shopping users' explicit/implicit feedback on the website, recommendation semantics and input deviation amount. Then, these features are used as input features of the BP neural network to classify and identify the priority of the final output web page. Finally, the web pages are sorted according to priority and recommended to users. This project uses book sales webpages as samples for experiments. The results show that this solution can quickly and accurately identify the webpages required by users.*

**Keywords:** E-commerce; webpage recommendation; semantic web mining; BP neural network.

## 1. INTRODUCTION

With the development of the Internet, e-commerce has become a major way for customers to purchase goods. Among them, effective web page recommendations can recommend the products customers need based on their shopping habits and improve shopping efficiency. For this reason, a variety of personalized recommendation systems[1] based on web mining[2] have been formed. For example, literature[3] proposes a web page recommendation algorithm based on semantic Web data mining technology. By analyzing the web pages visited by users, log data is used to mine the information and time of purchased goods to build a recommendation model. Literature[4] obtains search information based on user feedback, and constructs samples to train a reverse neural network to determine the importance of e-commerce web pages.

With the rapid development of the web industry, e-commerce businesses have become ubiquitous. In the absence of a web directory concept, users rely on search alerts to find suitable e-commerce web pages to purchase products. Usually search engines complete the matching process based on the statistical frequency and similarity between user search queries and candidate web pages. However, there are semantic problems with this matching method because user queries can be interpreted as different contents, resulting in incorrect retrieval results[5]. In addition, the web page ranking results provided by many well-known search engines are unreliable because they There is a commercial problem. The semantic web can be classified based on existing vocabulary and semantics, which greatly improves the accuracy. The literature[6] proposes to use semantic web to assist diagnosis and reduce the misdiagnosis rate. For promotion or advertising of some web pages, relevant business web pages are not sorted according to the user's search content, resulting in users not being able to accurately obtain the optimal web page for the desired product. One reason for this problem is that search alerts are generally designed There is no understanding of user needs. On the other hand, the reason is that the retrieval algorithm lacks error back propagation or feedback mechanism, causing the algorithm to often regard the web pages at the top as the web pages expected by the users[7].

In response to these problems, this paper proposes a web page recommendation algorithm to optimize the e-commerce web page ranking process, combines semantic web mining and BP neural network algorithms to handle the ranking problems of different types of web pages, and implements an intelligent meta-search engine to help users accurately obtain all the information they need. e-commerce web page required.

## 2. BP NEURAL NETWORK

BP neural network (Back-Propagation Neural Network, BPNN)[8] is a typical artificial neural network model. It has the advantages of simple structure and can effectively solve nonlinear function approximation problems. The basic idea of BP neural network is: if the output of the network is wrong, the weight of the network is adjusted so that the output of the network will develop in a smaller direction in the future, so that the output result is close to the expected value.

In this scheme, a three-layer BP neural network is used as the classifier for training and recognition. The network consists of an input layer, a hidden layer and an output layer, as shown in Figure 1.

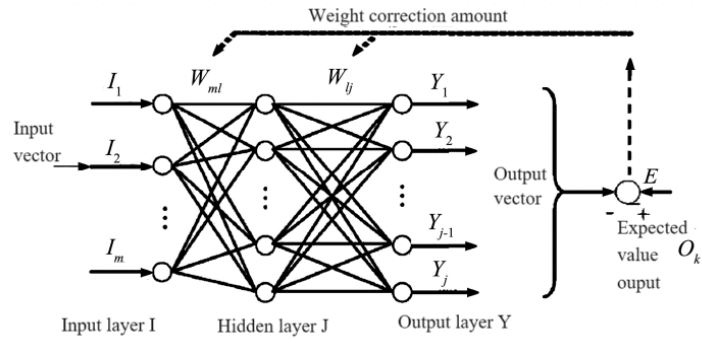

**Figure 1:** The BP neural network structure

This paper uses the traingd function as the training function and the tansig function as the transfer function between layers[9]. Through training, the weight matrix $W_{ji}$ and threshold matrix$\emptyset_j$from the input layer to the hidden layer are obtained; the hidden layer, the weight matrix $V_{ki}$ and the threshold matrix $\varphi_k$from the layer to the output layer.

In the BP neural network algorithm, the input $s_j$ of the $j$ neuron in the hidden layer is first calculated according to $W_{ji}$ and $\emptyset_j$ through Equation (1):

$$s_{j=}\sum_{i=0}^{I} W_{ji}\, x_j - \theta_j, j = 1,2, \dots, J \qquad (1)$$

The output $h_j$ of the j-th neuron in the hidden layer is obtained through transfer formula (2):

$$h_j = g(s_j) = \frac{2}{1+e^{-2s_j}} - 1, j = 1,2, \dots, J \qquad (2)$$

Then calculate the input $r_k$ of the output layer according to $V_{kj}$ and $\varphi_k$ through equation (3):

$$r_{k=}\sum_{j=1}^{J} V_{ji}\, h_j - \varphi_k, k = 1,2, \dots, K \qquad (3)$$

Obtain the output k y of the output layer through transfer equation (4):

$$y_k = g(r_k) = \frac{2}{1+e^{-2r_k}} - 1, k = 1,2, \dots, K \qquad (4)$$

Finally, find the maximum value among the recognition similarity values $y_k$, which is recognition results.

## 3. RECOMMENDED SOLUTIONS FOR E-COMMERCE WEB PAGES IN THIS ARTICLE

### 3.1 Web semantic mining

The semantic mining method is based on web mining. During the mining process, it integrates semantic-level "click stream" information, deeply mines various recommendation rules through logical reasoning, and generates recommended web pages. The recommendation method based on semantic mining has high accuracy and can effectively avoid the interference of irrelevant pages. To this end, this paper proposes to capture site semantics by integrating site content data to generate various recommendation rules, such as website content, web page dwell time, etc.

**3.2 Recommended solution framework for web pages in this project**

This project ranks web pages based on the user's search information, semantic web mining and BP neural network, recommends the web pages actually needed by the user, and increases the user's stay time on the website. The solution in this article adopts a modular design and is divided into preprocessing and dictionary implementation modules; time consumption priority module; content priority module; neural network priority module and semantic recommendation module.

In the preprocessing and dictionary implementation module, candidate web pages retrieved by search engines are preprocessed, incomplete input entries are removed, and data and stems are cleaned. This article uses the dictionary implementation module to input the website URL to implement the web log preprocessing process, implement user navigation analysis and a web page dictionary consisting only of words from candidate web pages, which are related to the length of words used by users to search for specific e-commerce products.

In the content priority module, web dictionaries and candidate web pages are used as input, and Web content mining technology is used to classify documents into an ordered structure, and then extract documents according to structural rules to extract knowledge from web document content or descriptions. Thereby retrieving the relevance of web pages, determining the priority of web pages, and removing and retrieving e-commerce website pages that are not related to service products.

In the time consumption priority module, the web page is taken as input and the candidate web page timestamp is used to determine the priority of the candidate web page. The timestamp is the statistical value of the time the previous user stayed on the same product web page. This module assigns higher priority to web pages where previous users spent more time.

In the semantic recommendation module, web pages are still used as input, the longest common subsequence algorithm is used to identify user session data of different semantic behavior files, and the ontology category is determined to avoid incorrect interpretations of user retrieval queries.

Finally, the overall priority of e-commerce web pages is determined through BP neural network. Content priority, time consumption priority, e-commerce users' explicit/implicit feedback on candidate websites, recommendation semantics and input bias amount are the 5 input features of the BP neural network. The neural network randomly assigns weights to all input nodes, compares the actual output of the network with a manually set threshold, and generates the difference between the actual output and the true value. The weight value of the input node is adaptively adjusted according to the feedback value of the error tolerance of the output to the input layer until the correct output is produced. Fine-tuning the weights in this way is important for supervised learning and can effectively improve the accuracy of prioritizing web pages related to user searches.

Since this project has 5 feature inputs, the input layer of the neural network consists of five neurons. Correspondingly, the output layer comprises a single neuron, and the output sample is the correlation between the web page represented by the input feature group and the user's preferences, which is the priority of the network. For the hidden layer, if the number of neurons is large, the classification accuracy will be improved to a certain extent, but it will also exponentially increase the transfer calculation from the input layer to the hidden layer and the hidden layer to the output layer, making it possible to learn and classify online. The time is too long. If the number of neurons is small, the network will have poor fault tolerance and more local minima. According to the experience of the existing literature, the optimal value for the number of neurons in the hidden layer is: the number of neurons in the input layer * the square root of the number of neurons in the output layer + L (L is an integer between 0 and 5), which is what this project implies the number of layer neurons can be selected from 3 to 8. Through multiple experimental analyses, this paper selected the number of neurons in the middle layer to be 4 under balanced consideration of classification performance and classification speed.

The BP neural network structure diagram used in this article is shown in Figure 2.

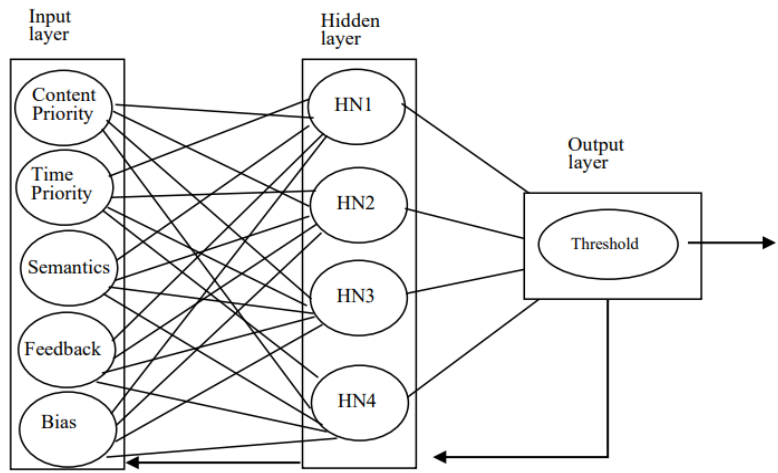

**Figure 2:** The BP neural network structure in this paper

The overall framework diagram of this article's solution is shown in Figure 3.

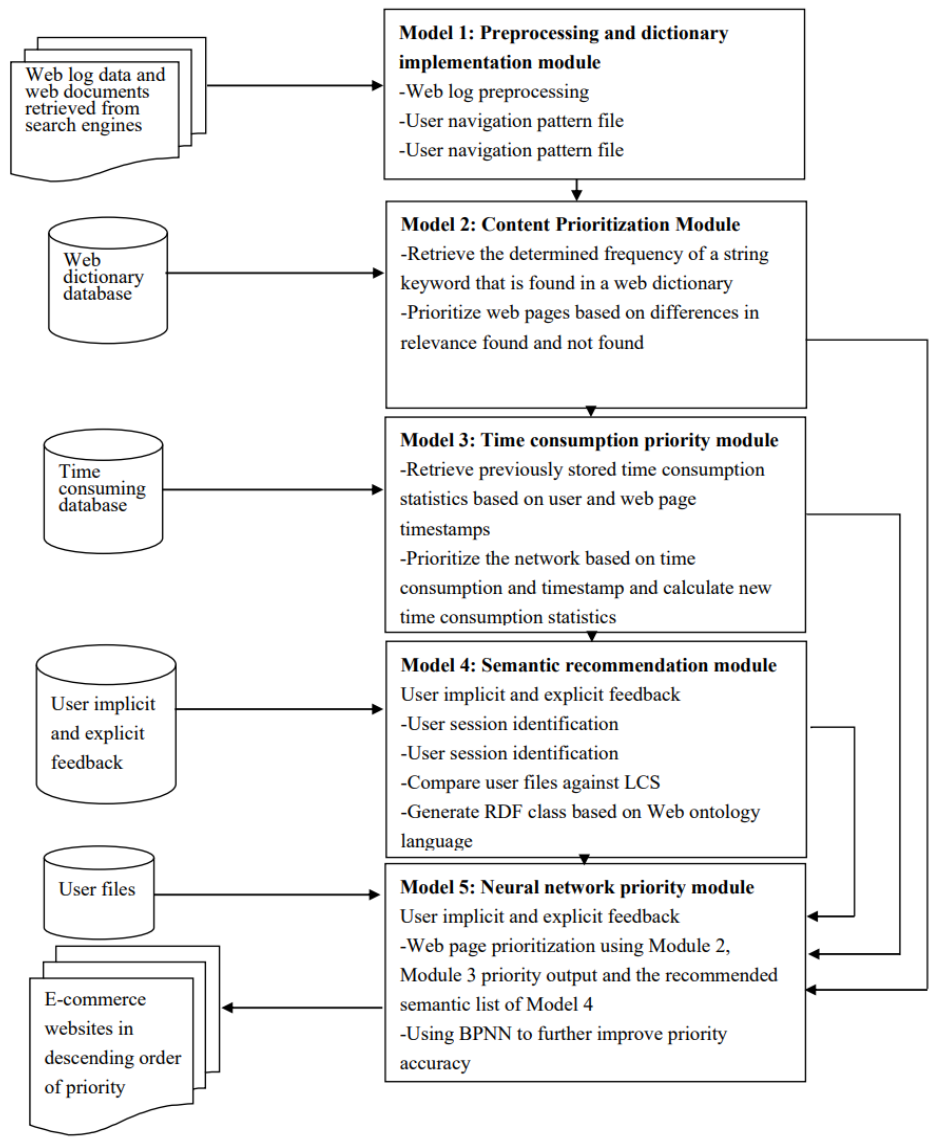

**Figure 3:** The framework of the proposed scheme

**3.3 This article's web page recommendation algorithm**

This section describes the specific algorithm of the e-commerce web page recommendation scheme proposed in this article based on semantic web mining and BP neural network. Among them, the variables and their meanings involved in the algorithm of this article are shown in Table 1.

**Table 1:** The variables in the proposed scheme

| Variable names | Meanings |
|---|---|
| $Si$ | The e-commerce product string retrieved by the user |
| Min | *The minimum length Si of any keyword in Si* |
| Max | The maximum length of any keyword in |
| $Wi$ | Search stage specific keywords |
| $Dp$ | Browsed e-commerce web documents |
| $WDp$ | Web dictionary corresponding to p Web document |
| $DW$ | Document word |
| $Tp$ | Average time spent by previous visitors |
| $Ts$ | Web page timestamp |
| Found | How often a keyword is found within among ranked e-commerce Web sites |
| Nfound | Frequency of keywords not found within $Si$ among ranked e-commerce Web sites |
| tan f | Linear activation function for training in neural networks |
| $WTi$ | The weight of the input neuron node |

As shown in Figure 3, the webpage recommendation solution in this article is divided into 5 modules. The algorithm flow of each module is as follows:

**Module 1:**

step 1: getting the retrieval string from the user.
step 2: removing stems from the search string.
step 3: recording the guided sequence pattern to the user file database.
step 4: using a search engine to retrieve web documents.
step 5: dividing the string into different distinct words $W_1, W_2,..., W_n$.
step 6: determining the minimum and maximum length of different words within the search statement.
min: $= Strlen(W_1),$ max: $= Strlen\ (W_1)$
for $i$ =2 to $n$ do
if min $> Strlen\ (W_i)$ then min: $= Strlen\ (W_i$ )
if max $< Strlen(W_i)$ then max: $= Strlen\ (W_i$ )
step 7: initializing Ti of each document to 0.
step 8: retrieving the time database based on the keywords entered by the user, and retrieve the same documents through the search engine when retrieving the Ti step.
step 9: preprocessing each web document $D_j$ in $WD_i$. There are only $DW_k$ words in $D_j$ and satisfy the condition $Strlen\ (DW_k) <=$ max.

**Module 2:**

step 10:   for $p$ =1 to $m$ do
initialization $found_p := 0, nfound_p := 0$
if $W_p$ is found in $WD_p$ then
$found_p := nfound_p + 1$
else $nfound_p := nfound_p + 1$
step 11: clearing web page $nfound_p > found_p$.

**Module 3:**

step 12: determining the timestamp of a web page $Ts$.

step 13: at the onset of the user session, determine the duration of the user session on the current web page $t_p$, and determined according to the following equation $T_p$:

if $T_p$ =0 then $T_p = t_p$

else $T_p = (t_p + T_p)/2$

step 14: if is lower and $T_p$ is higher, a higher weight is assigned to the web page.

step 15: updating time information with keywords, web addresses and $T_p$ tools.

**Module 4:**

step 16: user-directed sessions are identified through a comparative analysis between the user's search query and each entry within the user file database, conducted as follows:

$LCS[i,j] = 0$, if $i = 0$ or $j = 0$ OR

$LCS[i,j] = LCS[i-1, j-1] + 1$, if $i, j \neq 0$ and $S1_i = S2_j$

OR $LCS[i,j] = \max (LCS[i-1,j], LCS[i, j-1])$,

if $i, j > 0$ and $S1_i \neq S2_j$

step 17: generating categories using a web ontology language.

**Module 5:**

step 18: normalizing all priority inputs in modules 2, 3, and 4.

step 19: using the input and output data sets for training a BP neural network with a linear activation function, as described below:

$\{O\} = tan\emptyset\{I\}$

step 20: in the hidden layer and output layer, use the sigmoid function to predict the output. The S-shaped function is $\{O\} = [1/(1 + e^{-I})]$, and the summation function is $\sum(I_1 WT_1 + I_2 WT_2 + I_3 WT_3 + I_4 WT_4 + I_5 WT_5 + B)$ step 21: determining the error rate involved in weight adjustment for network neurons using the supervised BP neural network algorithm.

step 22: sorting web pages in descending order of web priority.

## 4. EXPERIMENTS AND ANALYSIS

### 4.1 Experimental setup

This article uses the ASP.NET framework to implement the recommendation algorithm proposed in this article, and compares the solution in this article with the literature[3]. Construct an experiment and set the website's sales products to be books[10]. The number of books in the database is 5,800, divided into education, humanities and social sciences, technology, 5 categories including literature and life. Each major category is subdivided into 6 subcategories. Taking science and technology as an example, it is divided into subcategories such as popular science, computers, architecture, medicine, agriculture and forestry, and science. After a period of time, more than 6,400 transaction events were collected. The set is divided into 5 parts, 1 part is the test set, and the other 4 parts are the training set. This article uses 2 commonly used performance indicators: precision and recall to evaluate web page recommendations program performance. The accuracy rate refers to the proportion of relevant documents in the recommendation results, that is, the ratio of recommended products calculated by the recommendation model to all products related to the recommended topic[11], the expression is as follows:

$$PRE = \frac{\sum_{u \in U} R(u) \cap T(u)}{\sum_{u \in U} R(u)} \tag{5}$$

Recall rate refers to the proportion of relevant documents included in the recommended results to all relevant documents in the entire collection. Recall rate measures whether all the recommended pages have been fully recommended[12]. The expression shown as below:

$$REC = \frac{\sum_{u \in U} R(u) \cap T(u)}{\sum_{u \in U} T(u)} \tag{6}$$

$R(u)$ is the recommendation list based on the recommendation algorithm on the test set, $T(u)$ is the recommendation list on the test set, U is the user set, and I is the product set.

## 4.2 Accuracy comparison

On the test set, the accuracy and recall rates of the scheme in this paper and the scheme in literature[3]are shown in Figure 4 and Figure 5 respectively.

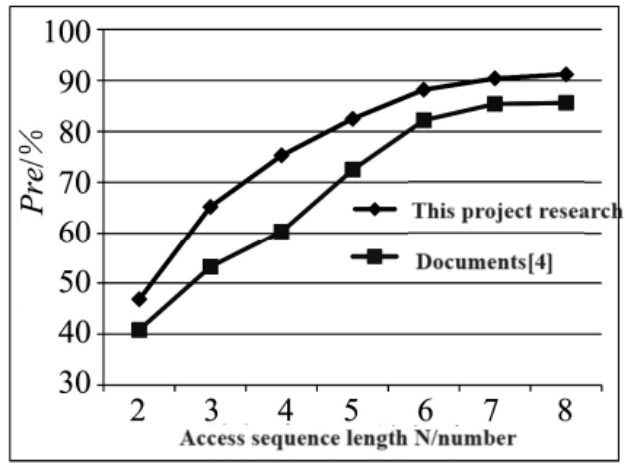

**Figure 4:** The precision rate of two schemes

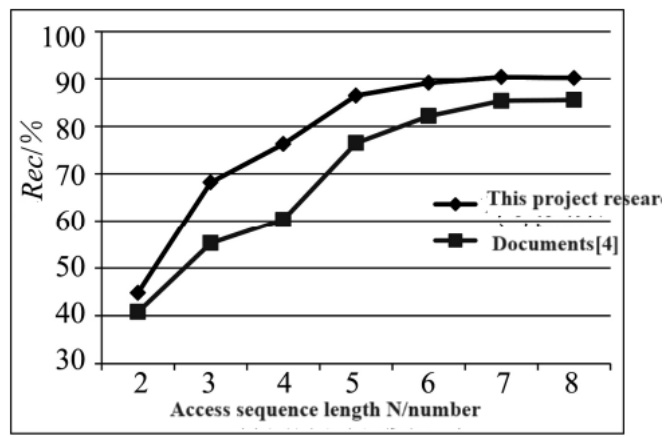

**Figure 5:** The recall rate of two schemes

Among them, taking the length of the user's access sequence as the abscissa, the access sequence is the process of the user's continuous access to the network. If a user's access sequence is <A, B, C>, it means that the user first accesses URLA, then accesses URLB, and finally Visit URLC. The longer the access sequence, the more user preference information it contains[13].

As can be seen from Figures 4 and 5, both the accuracy and recall of website recommendations increase as the access sequence length increases. This is because the access sequence length increases, making the features obtained by web data mining more representative of user preferences. However, when the access sequence length reaches 6, various performance metrics tend to stabilize. However, the scheme in this paper achieves better performance than that in literature[3]. When the sequence length is 8, the accuracy and recall rate of this scheme are 91.7% and 90.3% respectively, which are 6.2% and 4.9% higher than in literature[3] respectively. Experimental results show that this solution can better identify the web pages needed by users and is suitable for ranking and recommendation of e-commerce websites.

## 4.3 Time consumption comparison

The computational efficiency of this method is compared with the method in literature[3]. Among the above-mentioned collected web log files of more than 6,400 book transaction events, 10 K, 20 K, 50 K and 100 K log files were extracted respectively. The time consumed by the two algorithms in processing these files demonstrated in Table 2.

**Table 2:** Comparison of time consumption

| File size | 10K | 20K | 50K | 100K |
|---|---|---|---|---|
| Literature[3] | 100s | 567s | 1474s | 4657s |
| This project schema | 63s | 319s | 859s | 2260s |

It can be seen from Table 2 that when the file size is the same, the time consumed by the model in literature[3] is much higher than that of the solution in this paper, and the larger the file, the more obvious this trend is. Literature[3] analyzes the access records of a large number of web users to predict current user access behavior. However, since users' access to the web is related to various factors such as cultural background, interests, hobbies, and browsing purposes, and is not constrained by space and time, the web browsing behavior of each user also shows great differences. Using a semantic model describes the browsing behavior characteristics of all users and makes predictions, which inevitably has the disadvantages of low accuracy and high time and space complexity.

## 5. CONCLUSION

This paper proposes an e-commerce web page recommendation scheme based on semantic web mining and BP neural network. Recommend web page rankings based on user preferences, and may also help web designers optimize the structure of their company web sites. An experimental database of book sales webpages is constructed, and the recognition accuracy and recall rate of the webpage are used as performance indicators to compare the solution in this paper with the solution in the literature[3]. The results show that the solution in this paper has high recognition performance and can recommend what users need. web page information and has low computational complexity.

With the popularization of the application of deep learning algorithms, deep learning algorithms can be applied to image process[14], medical image analysis[15][16] and image classification[17] in the medical field. In the future, we can further explore the application of deep learning in the field of e-commerce. Neural network and deep learning technology can be used for face recognition[18] and can also predict future trends based on historical financial data[19]. Combined with machine learning, it can also be used for climate prediction [20]. Moreover, some algorithmic techniques are also used for classification and autonomous learning, etc.[21][22], combining these algorithms with user purchasing and browsing data to predict e-commerce sales trends is also a direction worth researching.